\documentclass{aastex63}
\usepackage{xcolor}
\usepackage{longtable}
\usepackage{amsmath,amssymb}
\usepackage{rotating}
\usepackage{amssymb}
\usepackage{natbib}

\makeatletter

\newcommand{\Rmnum}[1]{\expandafter\@slowromancap\romannumeral #1@}
\makeatother

\submitjournal{ApJ}

\shorttitle{Evolving spectral-timing properties of the Rapid Burster}
\shortauthors{Chen et al.}

\begin{document}


\title{New insight into the Rapid Burster by Insight-HXMT}

\author{Y. P. Chen\textsuperscript{*}}
\affil{Key Laboratory for Particle Astrophysics, Institute of High Energy Physics, Chinese Academy of Sciences, 19B Yuquan Road, Beijing 100049, China}

\author{S. Zhang\textsuperscript{*}}
\affil{Key Laboratory for Particle Astrophysics, Institute of High Energy Physics, Chinese Academy of Sciences, 19B Yuquan Road, Beijing 100049, China}

\author{S. N. Zhang}
\affil{Key Laboratory for Particle Astrophysics, Institute of High Energy Physics, Chinese Academy of Sciences, 19B Yuquan Road, Beijing 100049, China}
\affil{University of Chinese Academy of Sciences, Chinese Academy of Sciences, Beijing 100049, China}

\author{L. Ji}
\affil{Institut f{\"u}r Astronomie und Astrophysik, Kepler Center for Astro and Particle Physics, Eberhard Karls, Universit{\"a}t, Sand 1, D-72076 T{\"u}bingen, Germany}

\author{L. D. Kong}
\affil{Key Laboratory for Particle Astrophysics, Institute of High Energy Physics, Chinese Academy of Sciences, 19B Yuquan Road, Beijing 100049, China}
\affil{University of Chinese Academy of Sciences, Chinese Academy of Sciences, Beijing 100049, China}

\author{P. J. Wang}
\affil{Key Laboratory for Particle Astrophysics, Institute of High Energy Physics, Chinese Academy of Sciences, 19B Yuquan Road, Beijing 100049, China}
\affil{University of Chinese Academy of Sciences, Chinese Academy of Sciences, Beijing 100049, China}

\author{L. Tao}
\affil{Key Laboratory for Particle Astrophysics, Institute of High Energy Physics, Chinese Academy of Sciences, 19B Yuquan Road, Beijing 100049, China}

\author{M. Y. Ge}
\affil{Key Laboratory for Particle Astrophysics, Institute of High Energy Physics, Chinese Academy of Sciences, 19B Yuquan Road, Beijing 100049, China}

\author{C. Z. Liu}
\affil{Key Laboratory for Particle Astrophysics, Institute of High Energy Physics, Chinese Academy of Sciences, 19B Yuquan Road, Beijing 100049, China}

\author{F. J. Lu}
\affil{Key Laboratory for Particle Astrophysics, Institute of High Energy Physics, Chinese Academy of Sciences, 19B Yuquan Road, Beijing 100049, China}

\author{J. L. Qu}
\affil{Key Laboratory for Particle Astrophysics, Institute of High Energy Physics, Chinese Academy of Sciences, 19B Yuquan Road, Beijing 100049, China}
\affil{University of Chinese Academy of Sciences, Chinese Academy of Sciences, Beijing 100049, China}

\author{T. P. Li}
\affil{Key Laboratory for Particle Astrophysics, Institute of High Energy Physics, Chinese Academy of Sciences, 19B Yuquan Road, Beijing 100049, China}
\affil{Department of Astronomy, Tsinghua University, Beijing 100084, China}
\affil{University of Chinese Academy of Sciences, Chinese Academy of Sciences, Beijing 100049, China}

\author{Y. P. Xu}
\affil{Key Laboratory for Particle Astrophysics, Institute of High Energy Physics, Chinese Academy of Sciences, 19B Yuquan Road, Beijing 100049, China}


\author{X. L. Cao}
\affil{Key Laboratory for Particle Astrophysics, Institute of High Energy Physics, Chinese Academy of Sciences, 19B Yuquan Road, Beijing 100049, China}

\author{Y. Chen}
\affil{Key Laboratory for Particle Astrophysics, Institute of High Energy Physics, Chinese Academy of Sciences, 19B Yuquan Road, Beijing 100049, China}

\author{Q. C. Bu}
\affil{Key Laboratory for Particle Astrophysics, Institute of High Energy Physics, Chinese Academy of Sciences, 19B Yuquan Road, Beijing 100049, China}

\author{C. Cai}
\affil{Key Laboratory for Particle Astrophysics, Institute of High Energy Physics, Chinese Academy of Sciences, 19B Yuquan Road, Beijing 100049, China}

\author{Z. Chang}
\affil{Key Laboratory for Particle Astrophysics, Institute of High Energy Physics, Chinese Academy of Sciences, 19B Yuquan Road, Beijing 100049, China}

\author{G. Chen}
\affil{Key Laboratory for Particle Astrophysics, Institute of High Energy Physics, Chinese Academy of Sciences, 19B Yuquan Road, Beijing 100049, China}

\author{L. Chen}
\affil{Department of Astronomy, Beijing Normal University, Beijing 100088, China}

\author{T. X. Chen}
\affil{Key Laboratory for Particle Astrophysics, Institute of High Energy Physics, Chinese Academy of Sciences, 19B Yuquan Road, Beijing 100049, China}

\author{W. W. Cui}
\affil{Key Laboratory for Particle Astrophysics, Institute of High Energy Physics, Chinese Academy of Sciences, 19B Yuquan Road, Beijing 100049, China}

\author{Y. Y. Du}
\affil{Key Laboratory for Particle Astrophysics, Institute of High Energy Physics, Chinese Academy of Sciences, 19B Yuquan Road, Beijing 100049, China}

\author{G. H. Gao}
\affil{Key Laboratory for Particle Astrophysics, Institute of High Energy Physics, Chinese Academy of Sciences, 19B Yuquan Road, Beijing 100049, China}
\affil{University of Chinese Academy of Sciences, Chinese Academy of Sciences, Beijing 100049, China}

\author{H. Gao}
\affil{Key Laboratory for Particle Astrophysics, Institute of High Energy Physics, Chinese Academy of Sciences, 19B Yuquan Road, Beijing 100049, China}
\affil{University of Chinese Academy of Sciences, Chinese Academy of Sciences, Beijing 100049, China}

\author{M. Gao}
\affil{Key Laboratory for Particle Astrophysics, Institute of High Energy Physics, Chinese Academy of Sciences, 19B Yuquan Road, Beijing 100049, China}

\author{Y. D. Gu}
\affil{Key Laboratory for Particle Astrophysics, Institute of High Energy Physics, Chinese Academy of Sciences, 19B Yuquan Road, Beijing 100049, China}

\author{J. Guan}
\affil{Key Laboratory for Particle Astrophysics, Institute of High Energy Physics, Chinese Academy of Sciences, 19B Yuquan Road, Beijing 100049, China}

\author{C. C. Guo}
\affil{Key Laboratory for Particle Astrophysics, Institute of High Energy Physics, Chinese Academy of Sciences, 19B Yuquan Road, Beijing 100049, China}
\affil{University of Chinese Academy of Sciences, Chinese Academy of Sciences, Beijing 100049, China}

\author{D. W. Han}
\affil{Key Laboratory for Particle Astrophysics, Institute of High Energy Physics, Chinese Academy of Sciences, 19B Yuquan Road, Beijing 100049, China}

\author{Y. Huang}
\affil{Key Laboratory for Particle Astrophysics, Institute of High Energy Physics, Chinese Academy of Sciences, 19B Yuquan Road, Beijing 100049, China}

\author{J. Huo}
\affil{Key Laboratory for Particle Astrophysics, Institute of High Energy Physics, Chinese Academy of Sciences, 19B Yuquan Road, Beijing 100049, China}

\author{S. M. Jia}
\affil{Key Laboratory for Particle Astrophysics, Institute of High Energy Physics, Chinese Academy of Sciences, 19B Yuquan Road, Beijing 100049, China}

\author{W. C. Jiang}
\affil{Key Laboratory for Particle Astrophysics, Institute of High Energy Physics, Chinese Academy of Sciences, 19B Yuquan Road, Beijing 100049, China}

\author{J. Jin}
\affil{Key Laboratory for Particle Astrophysics, Institute of High Energy Physics, Chinese Academy of Sciences, 19B Yuquan Road, Beijing 100049, China}

\author{B. Li}
\affil{Key Laboratory for Particle Astrophysics, Institute of High Energy Physics, Chinese Academy of Sciences, 19B Yuquan Road, Beijing 100049, China}

\author{C. K. Li}
\affil{Key Laboratory for Particle Astrophysics, Institute of High Energy Physics, Chinese Academy of Sciences, 19B Yuquan Road, Beijing 100049, China}

\author{G. Li}
\affil{Key Laboratory for Particle Astrophysics, Institute of High Energy Physics, Chinese Academy of Sciences, 19B Yuquan Road, Beijing 100049, China}

\author{W. Li}
\affil{Key Laboratory for Particle Astrophysics, Institute of High Energy Physics, Chinese Academy of Sciences, 19B Yuquan Road, Beijing 100049, China}

\author{X. Li}
\affil{Key Laboratory for Particle Astrophysics, Institute of High Energy Physics, Chinese Academy of Sciences, 19B Yuquan Road, Beijing 100049, China}

\author{X. B. Li}
\affil{Key Laboratory for Particle Astrophysics, Institute of High Energy Physics, Chinese Academy of Sciences, 19B Yuquan Road, Beijing 100049, China}

\author{X. F. Li}
\affil{Key Laboratory for Particle Astrophysics, Institute of High Energy Physics, Chinese Academy of Sciences, 19B Yuquan Road, Beijing 100049, China}

\author{Z. W. Li}
\affil{Key Laboratory for Particle Astrophysics, Institute of High Energy Physics, Chinese Academy of Sciences, 19B Yuquan Road, Beijing 100049, China}

\author{X. H. Liang}
\affil{Key Laboratory for Particle Astrophysics, Institute of High Energy Physics, Chinese Academy of Sciences, 19B Yuquan Road, Beijing 100049, China}

\author{J. Y. Liao}
\affil{Key Laboratory for Particle Astrophysics, Institute of High Energy Physics, Chinese Academy of Sciences, 19B Yuquan Road, Beijing 100049, China}

\author{B. S. Liu}
\affil{Department of Astronomy, Tsinghua University, Beijing 100084, China}

\author{H. W. Liu}
\affil{Key Laboratory for Particle Astrophysics, Institute of High Energy Physics, Chinese Academy of Sciences, 19B Yuquan Road, Beijing 100049, China}

\author{H. X. Liu}
\affil{Key Laboratory for Particle Astrophysics, Institute of High Energy Physics, Chinese Academy of Sciences, 19B Yuquan Road, Beijing 100049, China}

\author{X. J. Liu}
\affil{Key Laboratory for Particle Astrophysics, Institute of High Energy Physics, Chinese Academy of Sciences, 19B Yuquan Road, Beijing 100049, China}

\author{X. F. Lu}
\affil{Key Laboratory for Particle Astrophysics, Institute of High Energy Physics, Chinese Academy of Sciences, 19B Yuquan Road, Beijing 100049, China}

\author{Q. Luo}
\affil{Key Laboratory for Particle Astrophysics, Institute of High Energy Physics, Chinese Academy of Sciences, 19B Yuquan Road, Beijing 100049, China}
\affil{University of Chinese Academy of Sciences, Chinese Academy of Sciences, Beijing 100049, China}

\author{T. Luo}
\affil{Key Laboratory for Particle Astrophysics, Institute of High Energy Physics, Chinese Academy of Sciences, 19B Yuquan Road, Beijing 100049, China}

\author{R. C. Ma}
\affil{Key Laboratory for Particle Astrophysics, Institute of High Energy Physics, Chinese Academy of Sciences, 19B Yuquan Road, Beijing 100049, China}

\author{X. Ma}
\affil{Key Laboratory for Particle Astrophysics, Institute of High Energy Physics, Chinese Academy of Sciences, 19B Yuquan Road, Beijing 100049, China}

\author{B. Meng}
\affil{Key Laboratory for Particle Astrophysics, Institute of High Energy Physics, Chinese Academy of Sciences, 19B Yuquan Road, Beijing 100049, China}

\author{Y. Nang}
\affil{Key Laboratory for Particle Astrophysics, Institute of High Energy Physics, Chinese Academy of Sciences, 19B Yuquan Road, Beijing 100049, China}
\affil{University of Chinese Academy of Sciences, Chinese Academy of Sciences, Beijing 100049, China}

\author{J. Y. Nie}
\affil{Key Laboratory for Particle Astrophysics, Institute of High Energy Physics, Chinese Academy of Sciences, 19B Yuquan Road, Beijing 100049, China}

\author{G. Ou}
\affil{Key Laboratory for Particle Astrophysics, Institute of High Energy Physics, Chinese Academy of Sciences, 19B Yuquan Road, Beijing 100049, China}

\author{N. Sai}
\affil{Key Laboratory for Particle Astrophysics, Institute of High Energy Physics, Chinese Academy of Sciences, 19B Yuquan Road, Beijing 100049, China}
\affil{University of Chinese Academy of Sciences, Chinese Academy of Sciences, Beijing 100049, China}

\author{L. M. Song}
\affil{Key Laboratory for Particle Astrophysics, Institute of High Energy Physics, Chinese Academy of Sciences, 19B Yuquan Road, Beijing 100049, China}
\affil{University of Chinese Academy of Sciences, Chinese Academy of Sciences, Beijing 100049, China}

\author{X. Y. Song}
\affil{Key Laboratory for Particle Astrophysics, Institute of High Energy Physics, Chinese Academy of Sciences, 19B Yuquan Road, Beijing 100049, China}

\author{L. Sun}
\affil{Key Laboratory for Particle Astrophysics, Institute of High Energy Physics, Chinese Academy of Sciences, 19B Yuquan Road, Beijing 100049, China}

\author{Y. Tan}
\affil{Key Laboratory for Particle Astrophysics, Institute of High Energy Physics, Chinese Academy of Sciences, 19B Yuquan Road, Beijing 100049, China}

\author{Y. L. Tuo}
\affil{Key Laboratory for Particle Astrophysics, Institute of High Energy Physics, Chinese Academy of Sciences, 19B Yuquan Road, Beijing 100049, China}
\affil{University of Chinese Academy of Sciences, Chinese Academy of Sciences, Beijing 100049, China}

\author{C. Wang}
\affil{Key Laboratory of Space Astronomy and Technology, National Astronomical Observatories, Chinese Academy of Sciences, Beijing 100012,China}
\affil{University of Chinese Academy of Sciences, Chinese Academy of Sciences, Beijing 100049, China}

\author{L. J. Wang}
\affil{Department of Astronomy, Beijing Normal University, Beijing 100088, China}

\author{W. S. Wang}
\affil{Key Laboratory for Particle Astrophysics, Institute of High Energy Physics, Chinese Academy of Sciences, 19B Yuquan Road, Beijing 100049, China}

\author{Y. S. Wang}
\affil{Key Laboratory for Particle Astrophysics, Institute of High Energy Physics, Chinese Academy of Sciences, 19B Yuquan Road, Beijing 100049, China}

\author{X. Y. Wen}
\affil{Key Laboratory for Particle Astrophysics, Institute of High Energy Physics, Chinese Academy of Sciences, 19B Yuquan Road, Beijing 100049, China}

\author{B. B. Wu}
\affil{Key Laboratory for Particle Astrophysics, Institute of High Energy Physics, Chinese Academy of Sciences, 19B Yuquan Road, Beijing 100049, China}

\author{B. Y. Wu}
\affil{Key Laboratory for Particle Astrophysics, Institute of High Energy Physics, Chinese Academy of Sciences, 19B Yuquan Road, Beijing 100049, China}
\affil{University of Chinese Academy of Sciences, Chinese Academy of Sciences, Beijing 100049, China}

\author{M. Wu}
\affil{Key Laboratory for Particle Astrophysics, Institute of High Energy Physics, Chinese Academy of Sciences, 19B Yuquan Road, Beijing 100049, China}

\author{G. C. Xiao}
\affil{Key Laboratory for Particle Astrophysics, Institute of High Energy Physics, Chinese Academy of Sciences, 19B Yuquan Road, Beijing 100049, China}
\affil{University of Chinese Academy of Sciences, Chinese Academy of Sciences, Beijing 100049, China}

\author{S. Xiao}
\affil{Key Laboratory for Particle Astrophysics, Institute of High Energy Physics, Chinese Academy of Sciences, 19B Yuquan Road, Beijing 100049, China}
\affil{University of Chinese Academy of Sciences, Chinese Academy of Sciences, Beijing 100049, China}

\author{S. L. Xiong}
\affil{Key Laboratory for Particle Astrophysics, Institute of High Energy Physics, Chinese Academy of Sciences, 19B Yuquan Road, Beijing 100049, China}

\author{R. J. Yang}
\affil{College of physics Sciences \& Technology, Hebei University, No. 180 Wusi Dong Road, Lian Chi District, Baoding City, Hebei Province 071002, China}

\author{S. Yang}
\affil{Key Laboratory for Particle Astrophysics, Institute of High Energy Physics, Chinese Academy of Sciences, 19B Yuquan Road, Beijing 100049, China}

\author{Y. J. Yang}
\affil{Key Laboratory for Particle Astrophysics, Institute of High Energy Physics, Chinese Academy of Sciences, 19B Yuquan Road, Beijing 100049, China}

\author{Y. J. Yang}
\affil{Key Laboratory for Particle Astrophysics, Institute of High Energy Physics, Chinese Academy of Sciences, 19B Yuquan Road, Beijing 100049, China}

\author{Q. B. Yi}
\affil{Key Laboratory for Particle Astrophysics, Institute of High Energy Physics, Chinese Academy of Sciences, 19B Yuquan Road, Beijing 100049, China}
\affil{University of Chinese Academy of Sciences, Chinese Academy of Sciences, Beijing 100049, China}

\author{Q. Q. Yin}
\affil{Key Laboratory for Particle Astrophysics, Institute of High Energy Physics, Chinese Academy of Sciences, 19B Yuquan Road, Beijing 100049, China}

\author{Y. You}
\affil{Key Laboratory for Particle Astrophysics, Institute of High Energy Physics, Chinese Academy of Sciences, 19B Yuquan Road, Beijing 100049, China}

\author{F. Zhang}
\affil{Key Laboratory for Particle Astrophysics, Institute of High Energy Physics, Chinese Academy of Sciences, 19B Yuquan Road, Beijing 100049, China}

\author{H. M. Zhang}
\affil{Key Laboratory for Particle Astrophysics, Institute of High Energy Physics, Chinese Academy of Sciences, 19B Yuquan Road, Beijing 100049, China}

\author{J. Zhang}
\affil{Key Laboratory for Particle Astrophysics, Institute of High Energy Physics, Chinese Academy of Sciences, 19B Yuquan Road, Beijing 100049, China}

\author{W. C. Zhang}
\affil{Key Laboratory for Particle Astrophysics, Institute of High Energy Physics, Chinese Academy of Sciences, 19B Yuquan Road, Beijing 100049, China}

\author{W. Zhang}
\affil{Key Laboratory for Particle Astrophysics, Institute of High Energy Physics, Chinese Academy of Sciences, 19B Yuquan Road, Beijing 100049, China}
\affil{University of Chinese Academy of Sciences, Chinese Academy of Sciences, Beijing 100049, China}

\author{Y. Zhang}
\affil{Key Laboratory for Particle Astrophysics, Institute of High Energy Physics, Chinese Academy of Sciences, 19B Yuquan Road, Beijing 100049, China}

\author{Y. F. Zhang}
\affil{Key Laboratory for Particle Astrophysics, Institute of High Energy Physics, Chinese Academy of Sciences, 19B Yuquan Road, Beijing 100049, China}

\author{Y. H. Zhang}
\affil{Key Laboratory for Particle Astrophysics, Institute of High Energy Physics, Chinese Academy of Sciences, 19B Yuquan Road, Beijing 100049, China}
\affil{University of Chinese Academy of Sciences, Chinese Academy of Sciences, Beijing 100049, China}

\author{H. S. Zhao}
\affil{Key Laboratory for Particle Astrophysics, Institute of High Energy Physics, Chinese Academy of Sciences, 19B Yuquan Road, Beijing 100049, China}

\author{X. F. Zhao}
\affil{Key Laboratory for Particle Astrophysics, Institute of High Energy Physics, Chinese Academy of Sciences, 19B Yuquan Road, Beijing 100049, China}
\affil{University of Chinese Academy of Sciences, Chinese Academy of Sciences, Beijing 100049, China}

\author{S. J. Zheng}
\affil{Key Laboratory for Particle Astrophysics, Institute of High Energy Physics, Chinese Academy of Sciences, 19B Yuquan Road, Beijing 100049, China}

\author{Y. G. Zheng}
\affil{Key Laboratory for Particle Astrophysics, Institute of High Energy Physics, Chinese Academy of Sciences, 19B Yuquan Road, Beijing 100049, China}
\affil{College of physics Sciences \& Technology, Hebei  University, No. 180 Wusi Dong Road, Lian Chi District, Baoding City, Hebei Province 071002, China}

\author{D. K. Zhou}
\affil{Key Laboratory for Particle Astrophysics, Institute of High Energy Physics, Chinese Academy of Sciences, 19B Yuquan Road, Beijing 100049, China}
\affil{University of Chinese Academy of Sciences, Chinese Academy of Sciences, Beijing 100049, China}




\begin{abstract}
We report the timing and spectral analyses upon of the type II X-ray bursts from the Rapid Burster (MXB 1730--335) observed by  Insight-HXMT and Swift/XRT.
{\bf By stacking the long-duration  bursts,  we find for the first time that the hard X-rays are lagging than the soft X-rays by 3 seconds. However, such a lag is not visible for the short-duration bursts, probably because of the poor statistics. For all bursts the energy spectrum is found  to be non-thermal, thanks to the broad band coverage of Insight-HXMT. These   findings put new insights  into the type-II bursts and require a temporally showing-up corona for possible interpretation.   }

\end{abstract}


\section{Introduction}
The Rapid Burster (hereafter RB), also named MXB~1730--335, is a transient low mass X-ray binary (LMXB),
 with outbursts lasting for   a month and recurring every   7 months \citep{Masetti2002}.
The RB was discovered in 1976 by \citet{Lewin1976}, located in  globular cluster  Liller 1, at a distance 7.9 kpc  \citep{Valenti2010},
 with the most  salient feature of  type-II X-ray bursts in quick succession with short intervals     10 s during certain periods.
For  distinguishing it from the nearby type-I X-ray burster, which produces type-I X-ray bursts with long intervals $\sim$ ks, it was named the Rapid Burster.
Type-I X-ray bursts, also named  thermonuclear bursts,
, are caused by unstable burning of the accreted
hydrogen/helium supplied by the low-mass companion star (for
reviews, see \citealp{Lewin1993,Cumming,Strohmayer,Galloway}).
Correspondingly,  the nearby (only 0.$^{\circ}$5 from the RB) bright type-I X-ray burster 4U~1728--34 was named the Slow Burster (hereafter the SB), which is generally difficult for non-imaging telescopes to distinguish between the two bursters.

 The Rapid Burster also produces type I bursts during its outbursts, with a recurrent timescales of hours \citep{Hoffman1978}. Different from the RB, the bursting pulsar (hereafter BP, also named GRO J1744--228) \citep{Fishman1995}, with both pulsations (a period of 467 ms) \citep{Kouveliotou1996} and type-II bursts
detected, shows no type-I bursts.
Type I bursts in the RB occur in outburst from 45 percent of the Eddington luminosity to almost
quiescence \citep{Bagnoli2013}. However, type II bursts only appear below a critical luminosity of
about 10 percent the Eddington luminosity \citep{Guerriero1999}. The behavior of the type II bursts
from the RB exhibits a series of burst patterns. It is believed that three modes \citep{Marshall1979, Guerriero1999} follow one another in a smooth transition to phenomenologically describe type
II bursting behaviour, i.e., mode 0, mode 1 and mode 2, with duration and complexity behavior
decreasing gradually. Based on population study of the type II bursts from the RB, rather than the
mode 0–2, \citet{Bagnoli2015} prefer adopting a simple classification between long and short bursts,
which reflects whether they interact with the persistent emission.

Type-II X-ray bursts vary in duration from
  130 ms to   11 min and interval from seconds to an hour  \citep{Bagnoli2015}, with  peak luminosities up to
the Eddington luminosity $L_{\rm Edd}$.
In contrast to thermonuclear flashes on a NS surface causing type-I bursts, the type-II bursts are attributed instead to episodic accretion instabilities  \citep{Lewin1993}.
The most important and unique property  of  the RB
is that the type-II bursts have a relaxation-oscillator behaviour: a clear correlation between the burst fluence, $E$ and the waiting time  to the next burst, $\Delta$$t$ (known as $E-\Delta$$t$ relation) \citep{Lewin1976}.
Naturally, an explanation to the relaxation-oscillator behaviour is that a reservoir of accretion mass is accumulated around the neutron star \citep{Lewin1993}.
The mass from the reservoir releases the gravitational energy  during the burst,
and the reservoir will be refilled until the next burst.
However, the mechanisms of type II bursts remain largely a puzzle in how the accretion material releases its potential energy  in the innermost  disk or near the NS surface.
Basically there is a  dichotomy between magnetospheric and nonmagnetospheric models for
the generation of the type II bursts from the RB, characterized by magnetospheric instability and disk instability, respectively
 (for reviews, see \citealp{Lewin1993,Spruit1993,Lewin1995,Bagnoli2015}).

The RB was observed by Hard X-ray Modulation Telescope (HXMT) with a total
exposure time of 60 ks in August 2017.
Accompanied  with  simultaneous Swift/XRT data, the large effective area and broad band energy provide us an excellent opportunity to explore the  burst behaviour  of  the RB.
This paper is organized as follows:
We present the Insight-HXMT and Swift/XRT observations used in this work of the RB and describe the data reduction procedures in Section 2.
The next section (Section 3) describes the HXD detection and the burst spectral evolution, by stacking dozens bursts. We give the possible understandings upon  the observed  phenomena  in the last section (Section 4).

\section{Observations and Data analysis}

\begin{figure}[t]
\centering
   \includegraphics[angle=0, scale=0.6]{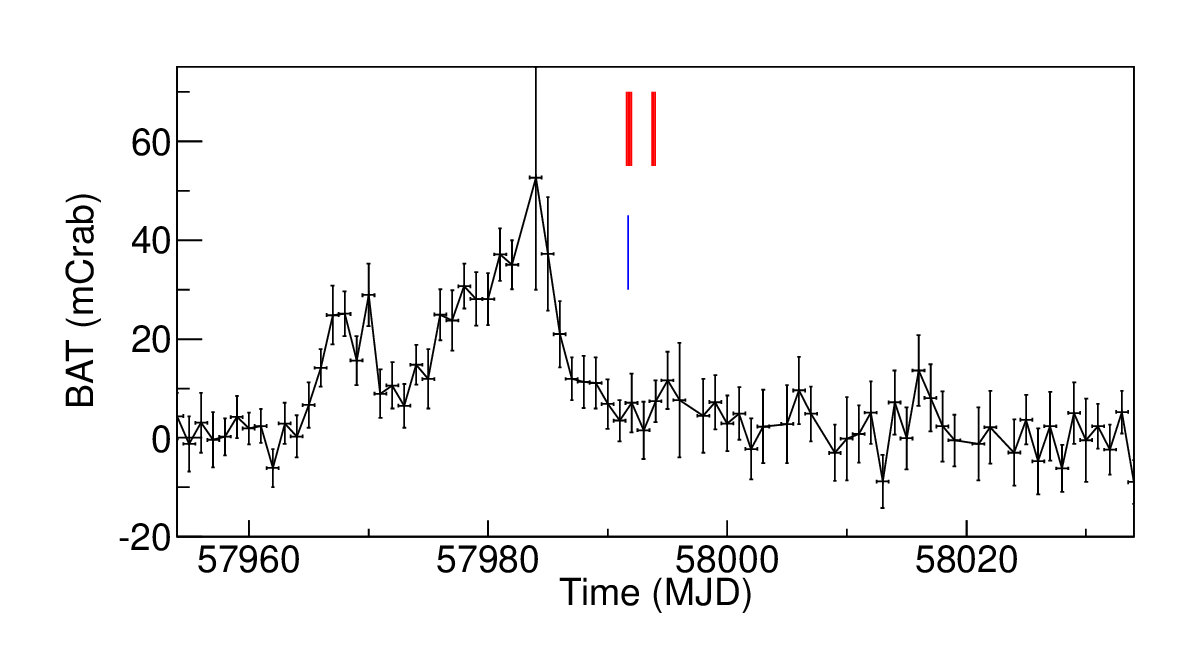}
 \caption{Daily lightcurves of the Rapid Burster by Swift/BAT during the outbursts in  2017 in 15--50 keV. The observations of Insight-HXMT and Swift/XRT  are indicated by vertical lines  by red and blue respectively.
 }
\label{lc_outburst}
\end{figure}

To explore the source state of the RB in Insight-HXMT observations, we take the Swift/BAT light curves. As shown in Fig. \ref{lc_outburst},
 the upper limit of the flux is 15 mCrab\footnote{https://swift.gsfc.nasa.gov/results/transients/},
 suggesting that the {\bf bursts are} located at the end of the outburst and evolves toward the quiescent state.

We use XSPEC v12.10.0 for the spectral analysis.  {\bf The interstellar
absorption model takes a column density fixed at  3.64 $\times10^{22}$~atoms/cm$^{2}$ \citep{vandenEijnden2017}, which is derived from a model consisting of a multicolor disk model (diskbb in XSPEC \citealp{Mitsuda1984}) and a disk reflection emission by fitting the persistent emission of the RB.}
The parameters of the X-ray spectra fitting model  are estimated with 68\% (1 $\sigma$) confidence level.

\subsection{Insight-HXMT}
 HXMT  (also
dubbed as Insight-HXMT, \citealp{Zhang2014},\citealp{Zhang2020}) excels in a broad energy band (1--250 keV)  and a large effective area in hard X-rays energy band.
It consists of three
slat-collimated instruments: the High Energy X-ray
Telescope (HE,  with a collection area 5103 ${\rm cm}^2$, 20--250 keV)\citep{Liu2020}, the Medium Energy X-ray Telescope
(ME, with  a collection area 952 ${\rm cm}^2$, 5--30 keV)\citep{Cao2020}, and the Low Energy X-ray Telescope (LE, with a collection area 384 ${\rm cm}^2$, 1--10 keV)\citep{Chen2020}.

HEASOFT version 6.22.1 and Insight-HXMT Data Analysis software
(HXMTDAS) v2.01 are used to analyze the data.
Only the small field of view (FoV) mode of LE and ME is used, for preventing   the contaminations from near-by sources and the bright earth.
The good time interval is filtered
 with the  recommended  criteria in HXMT Data Reduction Guide Version 2.01 given by the Insight-HXMT team.


Among the outburst of the Rapid Burster (RB) in 2017, there are 7 obsids on  the Slow Burster (SB), covering the time span between August 26th and 28th.
Due to the large FoV of Insight-HXMT, the RB was also detected in the FoV of these observations.
We performed a visual inspection of the light curves of the obsids, which results in
totally  $\sim$ 200 bursts, including a type-I X-ray burst.
All the type II bursts have a similar profile, seconds duration and tens seconds {\bf recurrence} time, indicating that  all of them are during mode-2 bursting phases towards the end of the outburst.

 LE and ME each contains 3   independent detector boxes, each with elongated rectangle FoV and different orientations.
 In these observations, the SB is located at the center of the FoVs and the RB is off center.
  If the bursts are from the SB, the fluxes of the 3 boxes should be roughly same, otherwise from RB.
The average flux of the type-I X-ray burst by the three LE boxes in 1--10 keV is 68.0$\pm$1.5, 70.5$\pm$1.6 and 68.2$\pm$1.5, respectively.
The roughly same counts rates in three boxes of LE or ME denote that the type-I burst should originate from the SB.

From these type-II X-ray bursts, we choose 159 bursts to analyse, which satisfies the  good-time-interval selection criteria of LE and ME  simultaneously,
 as shown in Table \ref{table_hxmt}. 
Among them, the bursts of the first three obsids have longer time intervals 150--280 s. 
 For the last three obsids, the bursts have shorter time intervals   30--50 s. 
For the bursts in the 4th obsid, both of the interval and duration vary significantly.
Additionally, there is a type-I X-ray burst in this obsid.
Based on the main characteristics of the bursts, we divide the bursts into two groups to analysis: the long bursts in the first 3 obsids (TOO1) and the short burst in the last 3 obsids (TOO2).

In the light curve and spectral analysis, we use the time when the burst reached its peak of ME as a reference for
producing the light curve and spectrum of the each burst.
With respect to the reference time, for the bursts  of TOO1 and TOO2, we split each burst into a sequence of 1 s  slices  after the burst onset, and stack the light curve slices in chronological order.
For the  burst spectra of TOO1 and TOO2, we split each burst into a sequence of 2 s and 1 s slices  after the burst onset, respectively.
Because of high background   and low flux of the RB detected by HE, only the LE and ME data are used to explore the evolution of the bursts.
We use the ftool  \emph{addspec} to stack the  spectral slices in chronological order, and fit the  time-ordered stacked spectra.

 Those time periods without burst are adopted as background, which includes instrumental background,  cosmic diffuse background
and persistent/non-burst flux of the RB and the SB, for  investigating  the burst spectrum evolution.


\subsection{Swift/XRT}

Within the same day when Insight-HXMT detected bursts from the RB, Swift/XRT also observed the source.
The XRT observation was performed  {\bf 2.8} hours after the Insight-HXMT observations, and cover a period of MJD 57991.69025 (UTC 2017-08-26 16:32:48)--57991.95835 (UTC 2017-08-26 22:58:56).
The OBSID is 00031360136 with an exposure    2 ks.
 The standard data process procedure of the timing mode is carried out  using the ftools
package with standard procedures (xrtpipeline version 0.13.5) \footnote{https://swift.gsfc.nasa.gov/analysis/}.  To avoid spectral pile-up distortions, the spectra of source and background  are extracted from  a source region
and a source-free background region with the grade 0.

\begin{figure}[t]
\centering
   \includegraphics[angle=0, scale=0.6]{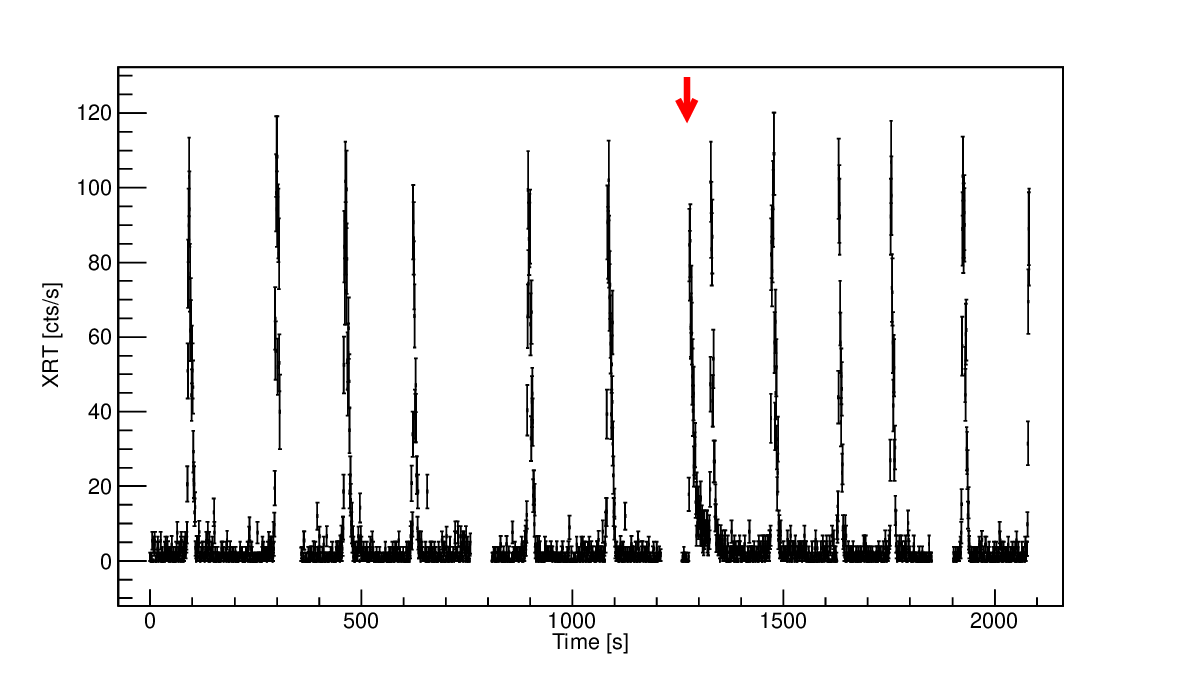}
 \caption{Sandwiched light curve of the RB during its 2017 outburst observed by Swift/XRT, for the complete Swift/XRT bandpass at 1 s resolution. Data gaps (50 s) mark jumps in time when there were no GTIs. The red arrow indicates  a type I X-ray burst from the RB.
}
\label{lc_xrt}
\end{figure}

A first inspection of the light curves shows strong variability with several bursts, and  the maximum count rate is   110 cts/s.
From the shapes of the bursts,  as shown in Fig. \ref{lc_xrt}, totally, 12 type-II bursts and 1 type-I burst are detected by Swift/XRT.  The persistent/non-burst emission is stable in this observation, remaining at  a count rate  2 cts/s.


\section{Results}
\subsection{Persistent emission}

\begin{figure}[t]
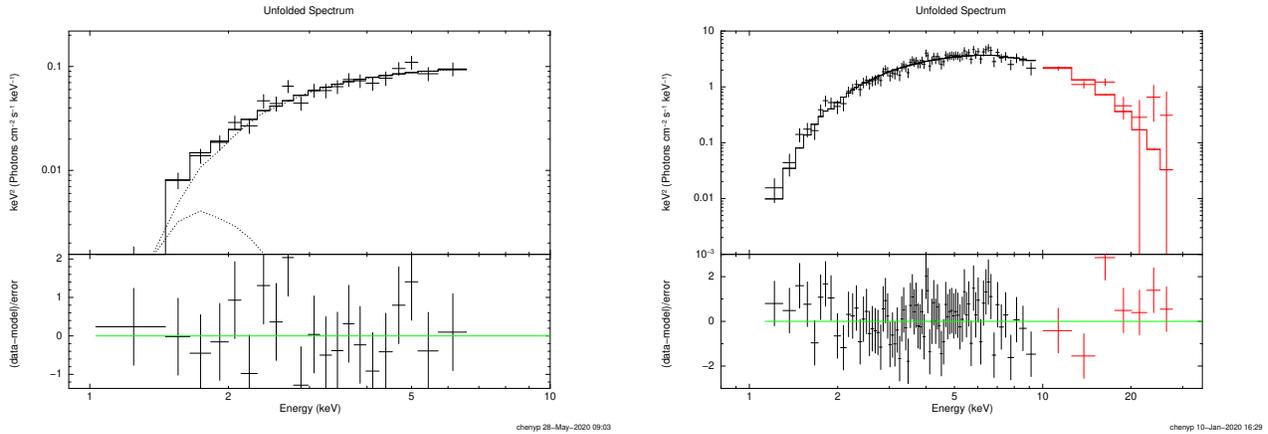

\centering
   \includegraphics[angle=270, scale=0.3]{xrt_persist.eps}
      \includegraphics[angle=270, scale=0.3]{fit_3_1_100.eps}
 \caption{The unfolded spectra of the  persistent/non-burst emission detected by Swift/XRT and the spectra of the TOO1 burst at the peak flux observed by HXMT/LE (black) and HXMT/ME (red) are given in left and right panel respectively.}
\label{spe_xrt}
\end{figure}

We fit the persistent spectrum observed by Swift/XRT, as show in Fig. \ref{spe_xrt},
with a model consisting of an interstellar absorption TBabs \citep{Wilms2000}, a power law and a  diskbb.
We find an acceptable  fit:$\chi^{2}_{\upsilon}$=0.83 ($\nu$ 17; Fig. \ref{spe_xrt}).
The unabsorbed bolometric flux in 1--10 keV is   $4.9_{-1.2}^{+0.6}\times10^{-10}~{\rm erg}~{\rm cm}^{-2}~{\rm s}^{-1}$,
  corresponding to  a {\bf luminosity} of $3.6\pm0.4\times10^{36}~{\rm erg}$~${\rm s}^{-1}$ at a distance of 7.9 kpc.



We explore a number of different fits, e.g., TBabs*(bbody+powerlaw),  and find that the blackbody model yields a good fit, with similar temperature, photo index and $\chi^{2}_{\upsilon}$ as derived from the diskbb model.
The energy range of the spectrum prevents us adopting more complex components to replace the power law model, e.g., a cutoff-powerlaw or a disk-reflection model.

\subsection{Timing properties}

\begin{figure}[t]
\centering
   \includegraphics[angle=0, scale=0.6]{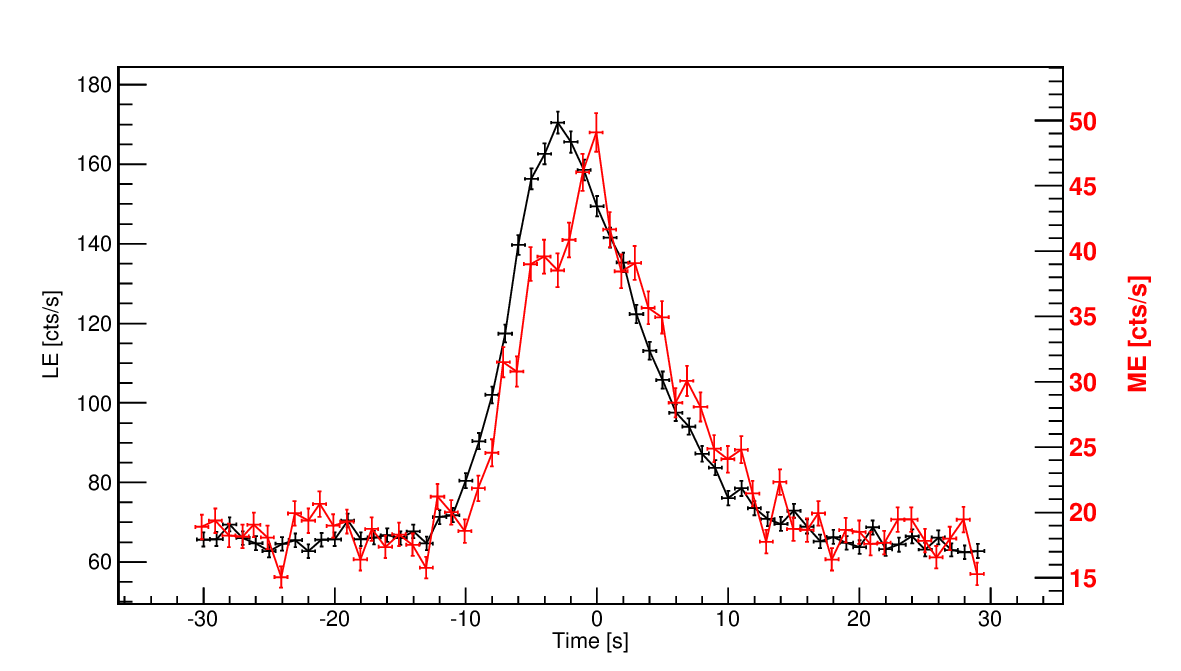}
   \includegraphics[angle=0, scale=0.6]{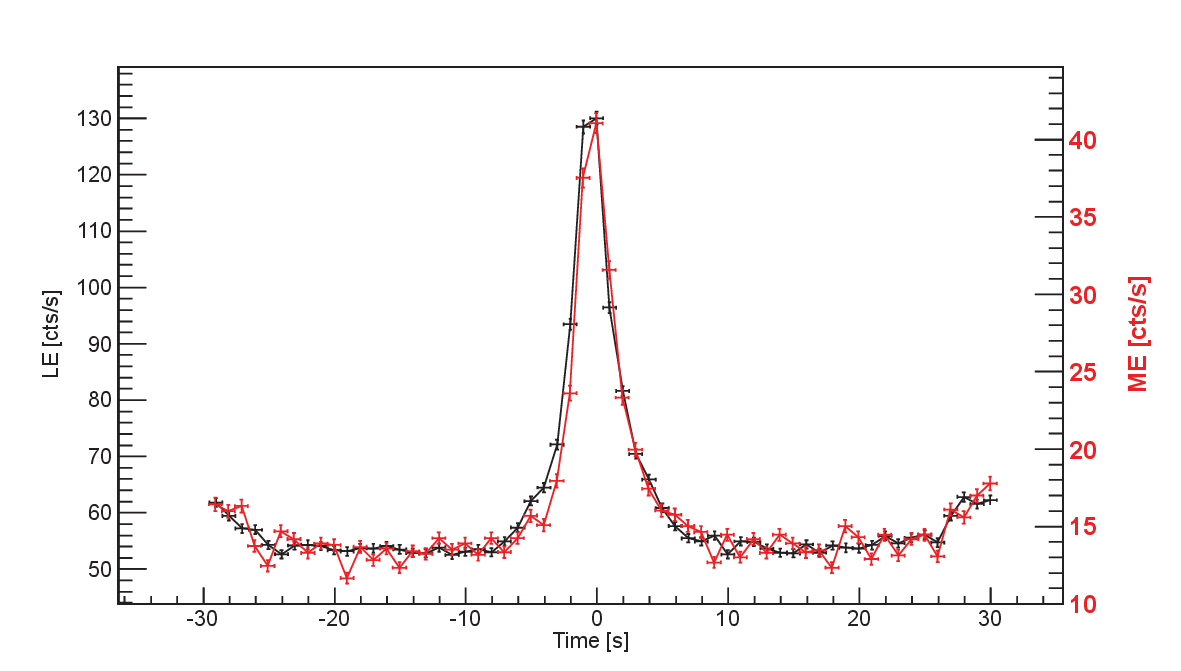}
 \caption{The stacked Insight-HXMT/LE and Insight-HXMT/ME light curves of the RB at 1--10 keV and 10--20 keV for TOO1 (the top panel) and TOO2 (the below panel).
 We note that in the upper panel, the hard X-ray emission (red) lags the soft ones, i.e.,  the HXD phenomenon.
 }
\label{lc_hxmt_burst_1}
\end{figure}

Since   low count rate   of LE\&ME  makes   difficult to analyze individual burst, we adopt a method that stacking bursts light curves and spectra.
For TOO1 and TOO2, the lightcurves of 23 and 117 bursts are stacked respectively.
As shown in Fig. \ref{lc_hxmt_burst_1},   the stacked light  curves of the bursts from the two groups have a similar trends in rise and decay, in the energy range 1--10 and 10--20 keV, respectively.
 The profile of the burst of TOO1 and TOO2 is  a Gaussian-like profile with a similar rising and decay time.
However, different from TOO2, the peak   of LE and ME in TOO1 occurs not  simultaneously, i.e.,
there is a   3 s delay of hard X-ray emission with respect to  the soft X-ray emission.
  For the sake of clarity, the soft X-ray flux and the  hard X-ray flux are scaled into a same panel.
 However, the  low count rate and short burst duration prevent us from re-bining the lightcurve of the TOO2 bursts with smaller time bin.
As shown in Fig. \ref{lc_hxmt_burst_1},   the burst width is roughly same  in the soft and hard X-ray band.


\subsection{Borad band spectrum}

The first attempt to model the  spectrum during the  burst evolution is an absorbed blackbody.
However, the result of the  $\chi^{2}_{\upsilon}$ (with $\nu$ 43--83) are mostly larger than 2, whether let the $N_{\rm H}$ free or not during spectral fitting.

An alternative model for the type-II X-ray burst is an absorbed Comptonization model (comptt in XSPEC) \citep{Titarchuk1994}.
This model yields a reasonable fit with  $\chi^{2}_{\upsilon}$ $\sim$1(with $\nu$ 43--83).
First, we let the seed photon energy $T_{\rm seed}$ and optical depth $\tau$ free during fitting.
In the course of the burst evolution, the fit results indicate the $T_{\rm seed}$ always around 0.2 keV, both for TOO1 and TOO2;
$\tau$ is 8.7 and 10 for TOO1 and TOO2, respectively.
Hence, we fixed $T_{\rm seed}$ and $\tau$ to explore the burst evolution, resulting variations are qualitatively very similar
to those measured with the free values, as shown in Fig. \ref{spe_xrt}.

The improved spectral fitting results in the electron temperature $kT_{\rm e}$ 2--3 keV and  the reduced $\chi^{2}_{\upsilon} $~0.7--1.4 with the $\nu$ 43-83.
As shown in Fig. \ref{burst_evolution}, both for TOO1 and TOO2, the profile of   $N_{\rm comptt}$ is similar, peaking at the center of the burst.
A similar trend is  also found for $kT_{\rm e}$ in TOO2.
For  TOO1  with the HXD phenomenon,  $kT_{\rm e}$  shows a  substantial increase from  2 to   3 keV.   
We combine  the tail of the burst spectra and find that   $kT_{\rm e}$  returns to 2 keV as measured at the initial phase.

 The burst unabsorbed bolometric peak fluxes in 1--30 keV  of TOO1 and TOO2 are obtained as $12.3 \pm 0.3$ and $ 7.0\pm 0.1$  $\times 10^{-9}~{\rm erg}~{\rm cm}^{-2}~{\rm s}^{-1}$,
  corresponding to  $9.2\pm 0.2$   and  $5.2 \pm 0.1 \times10^{37}$ erg s$^{-1}$ at a distance of 7.9 kpc, respectively.

{\bf We notice that} the spectral evolution of the heartbeat mode variability  in GRS~1915+105 is well described by a diskbb model in the course of the burst \citep{Taam1997,Paul1998},  with the disk $kT$ changed from 1.1 to  2 keV \citep{Mineo2012}. We analysed the burst spectra of the RB with the same model, and yield an acceptable reduced $\chi^{2}$. However, the disk $kT$ is up to 3 keV, which is too high for a NS X-ray binary.





\section{Discussion and Summary}

\begin{figure}[t]
\centering
  \includegraphics[angle=0, scale=0.3]{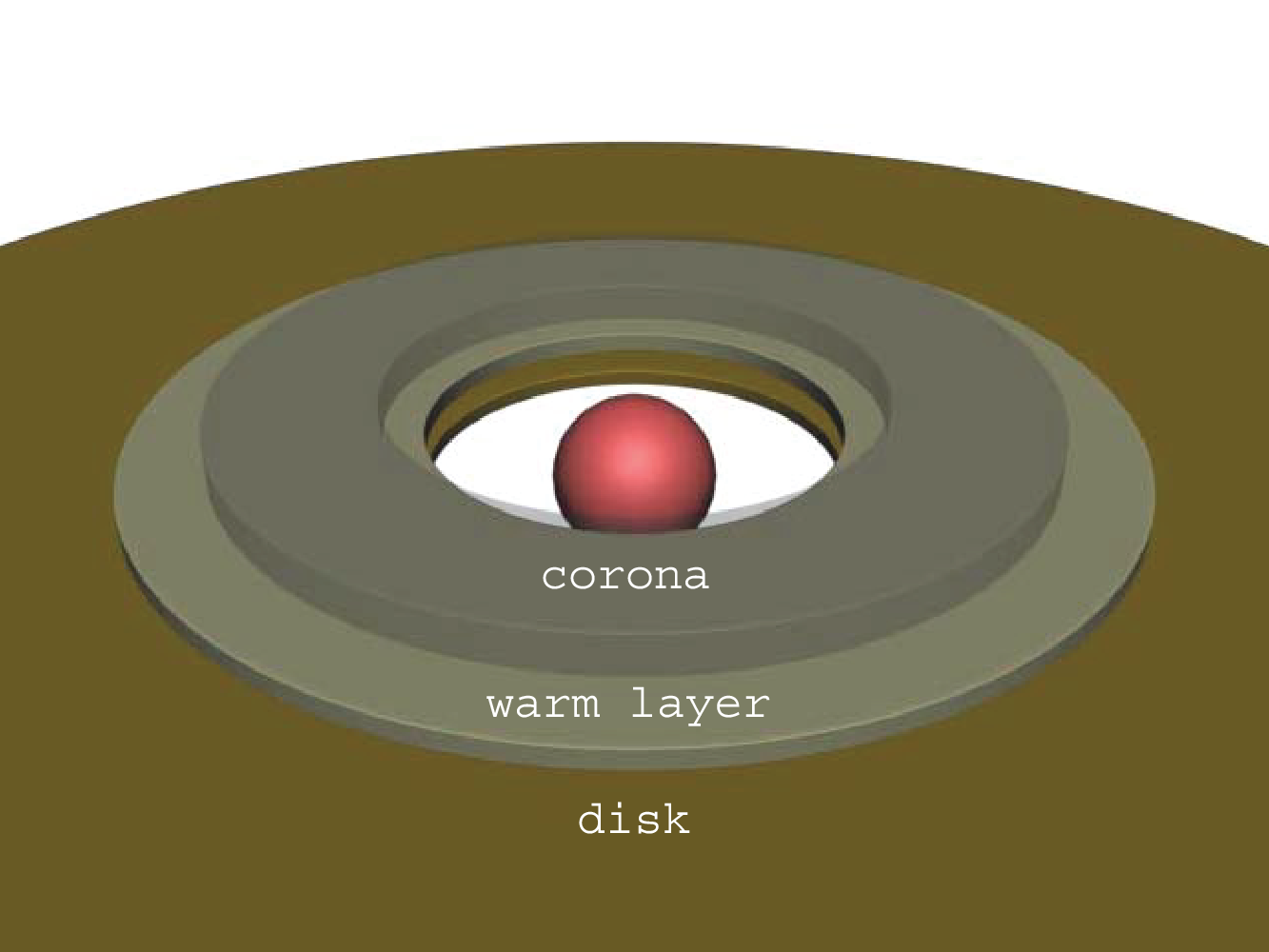}
   \caption{
   Illustration of the central region of an NS XRB during a  type-II X-ray burst,  in which the warm layer locates above the disk.
   }
\label{illustraction}
\end{figure}

{\bf 
We have analyzed the decay phase of the outburst experienced by the RB during 2017 observed by Insight-HXMT in a broad energy band. By stacking the type-II bursts with recurrence time of 100--300 seconds, we find for the first time a time delay of about 3 seconds at hard X-rays. 
The energy spectrum of the burst can be denoted by a Comptonization model, for which the temperature of the seed photons, the optical depth, the temperature of the hot plasma are derived  with values of 0.2 keV, 9--10 and 2--3 keV, respectively. These findings put more constraints upon the models at work for the RB.

The tight relation between the burst strength and the waiting time \citep{Lewin1976} leads to a series of models in history proposed via focusing on the sudden release of the accretion power in the disk. Type-II burst can be caused by the instability of the disk on viscous  \citep{Lightman1974} or thermal time scales  \citep{Shakura1976}, the magnetic barrier \citep{D2010,D2012} or the magnetic reconnection on the disk \citep{Jiang2014}.

However, all models have their own shortages in covering all the features observed in the RB \citep{Lewin1993}. The possible breakthrough to discriminate different models may come from the joint observations carried out in 2015 by XMM, Swift/XRT and Nustar  \citep{vandenEijnden2017}. With this campaign, both the continuous and the mode-0 burst spectra were well measured in a rather broad energy band of 1--30 keV. It turns out that, the inner radius keeps staying at round 41 $R_{\rm g}$  ($R_{\rm g}$=2G$M$/c$^{2}$)  for both burst and pre-burst emissions, and the corona temperature of the pre-burst is 7.19 keV, which is significantly larger than 2.42 keV measured for the non-thermal temperature of the burst. Comparison for the corona temperatures between burst and pre-burst are always challenged by the contamination source 4U 1728--34 which is located only 0.$^{\circ}$5 away from the RB.   We note that less change of the burst temperature with respect to the pre-burst was reported in ACSA \citep{Mahasena2003} and BeppoSAX \citep{Masetti2000}, which is probably due to limited bandwidth (i.e. below 10 keV) of the former and the relatively poorer statistics of the latter which does not allow to well constrain the parameters in comptt model. If the burst is caused by the temporary accretion on the disk, via either disk instability or the magnetic barrier, the inner disk radius should evolve to be smaller than that of the pre-burst. Regarding to the temperature difference of the corona between burst and pre-burst, the hot corona wherever is located, either in boundary layer or around the disk, can in principle be cooled by the seed photons of the burst shower. However, in history there are no hints for detecting such a cooling procedure in phase-resolved burst spectrum and all showed a constant corona temperature around 2 keV as derived from observations of ASCA \citep{Mahasena2003} and RXTE \citep{Bagnoli2015}. Here Insight-HXMT provides an additional clue for supporting no cooling effect for the corona during the burst. These results hence support a temporary show-up of an additional corona with temperature around 2 keV, via a mechanism of e.g. magnetic reconnection on the disk but rather than instability of the disk accretion.

During the decay phase of the 2017 outburst, Insight-HXMT observed for the type-II burst that the flux in 10--20 keV lags 3 seconds than that in 1--10 keV. Time delay of a few seconds was also reported in the heart-beat mode of micro-quasar GRS 1915+105 \citep{Janiuk2005,Neilsen2011,Massa2013}. \citet{Bagnoli2015} found in the RB the heart beat oscillation similar to GRS~1915+105 from  RXTE observations. However, a further investigation by \citet{Maselli2018} revealed no time lag between soft and hard X-rays during the two RXTE observations where the heart beat of the RB was discovered. 
Modellings of such hard X-ray delay are relatively rare so far and among a few trials are those from \citet{Janiuk2005,Neilsen2011,Massa2013}. 
\citet{Janiuk2005} considered radiation-pressure instability \citep{Lightman1974}  in the inner disk parts and showed that the hard X-ray delays can be caused by the time needed for the corona to adjust to the changing conditions in the underlying disk. In their theory, due to the instability of the radiation pressure mass of the inner accretion flow can be exchanged between the disk and the corona via heating and evaporation,   which forms a “density wave” around 20--30 $R_{\rm g}$ to cause a   loop.
Accordingly, the density of the corona evolves with time, which can end up at time scale of 1 second.
  Here we render the similar scenario for the hard X-ray time lag observed in type-II burst of the RB. 
We consider an instability caused by magnetic reconnection on the disk. Such a magnetosphere storm resulted from disk magnetic reconnection was initially proposed by \citet{Davidson1982}.  Magnetic reconnection can supply power to heat the disk and hence play a role in producing a temporary corona on top of the disk via extracting the accretion materials from the disk. Such a corona formation procedure can in principle result in hard X-ray delay at time scale of second in a way induced in \citet{Janiuk2005}. Regarding to the corona temperature, \citet{Zhang2000} showed that, the magnetic reconnection in the disk can produce a so-called warm layer, i.e. a corona with low temperature covering the cold disk, as shown in Fig. \ref{illustraction}. Therefore, we speculate that the temporary warm layer born out of the disk magnetic reconnection may correspond to the type-II burst observed in rapid burster.

In \citet{Zhang2000} the warm layer is supposed to be optical thick, which is consistent with the optical depth as measured during the type-II burst. We calculate the flux of the disk (diskbb) and warm  layer (comptt) in units of photon cm$^{-2}$ s$^{-1}$  with those model parameters derived in broad band spectral fitting with Insight-HXMT, and find that the two values are roughly same. 
This calculation indicates that nearly all of the disk emission are scattered to higher energy by the warm layer, and hence fits the scenario that part of disk is fully covered by the temporarily produced  corona. We notice that in Fig. \ref{burst_evolution} the corona temperature keeps increasing for long burst in TOO1 but turns over for short burst in TOO2. If the corona temperature is assumed to be the virial temperature \citep{Janiuk2005}, then during burst the temporary corona moves inward would lead to an increase in temperature. For the short burst in TOO2, gradual cease of the power supply from the disk magnetic reconnection in the decay phase of the burst can induce a drop in corona temperature.


In summary, the broad band observations of the rapid burster by Insight-HXMT reveal several interesting temporal and spectral properties of the type-II bursts born out of the outburst decay in 2017. These findings may be accounted for via introducing a temporary warm layer produced with disk magnetic reconnection. Further investigations upon the RB observed and the bursting pulsar by Insight-HXMT can definitely provide more clues to pin down this puzzle.
}

\acknowledgements
 This work made use of data from the Insight-HXMT mission, a project funded by China National Space Administration (CNSA) and the Chinese Academy of Sciences (CAS).
This work is supported by the National Key R\&D Program of China (2016YFA0400800) and the National Natural Science Foundation of China under grants  U1838201, U1938101, U1838202, 11473027, 11733009, U1838115.

\bibliographystyle{plainnat}

\clearpage

\begin{table}[ptbptbptb]
\begin{center}
\caption{The type-II X-ray bursts  detected by Insight-HXMT during 2017 outburst from the RB. }
 \label{table_hxmt}
\begin{tabular}{cccccccccccc}
\hline
OBSID &  Start Time & Elapsed Time (s) &	GTI (s) &	$n_{\rm burst}$	& $t_{\rm {\bf average} interval}({\rm s})$  \\\hline
P010130300101-20170826-01-01  & 2017-08-26T13:45:21& 11760	&1980	&7	&282.9 \\
P010130300102-20170826-01-01 & 2017-08-26T17:16:56 & 6600	&1635	&9	&181.7 \\
P010130300103-20170826-01-01 & 2017-08-26T20:27:50 &5820	&1136	&7	&162.3 \\
P010130300104-20170826-01-01 & 2017-08-26T23:38:44&11880	&1718	&19	&90.4 \\
P010130300201-20170828-01-01 & 2017-08-28T16:43:29&6120 &1620	&31	&52.3 \\
P010130300202-20170828-01-01 & 2017-08-28T20:11:30&5880 &1086	&31	&35.0 \\
P010130300203-20170828-01-01 & 2017-08-28T23:22:25&12120 &1785	&55	&32.5\\\hline
\hline
\end{tabular}
\end{center}
\end{table}

\begin{table}[ptbptbptb]
\begin{center}
\caption{The spectral fit results of the persistent emission by Swift/XRT. }
 \label{table_xrt}
\begin{tabular}{cccccccccccc}
\hline
$T_{\rm in}$ (keV)      & $R_{\rm in}$$^{*\mathrm{a}}$ (10$^{2}$km) & $\Gamma$ & $N^{*\mathrm{b}}$ & $\chi^{2}_{\upsilon}$($\nu$)	 \\\hline
$0.21_{-0.09}^{+0.16}$   & $1.65_{-0.81}^{+149}$                  & $1.97_{-0.15}^{+0.18}$ & $9.57_{-3.48}^{+2.29}$ & 0.83(17)
\\\hline
\end{tabular}
\end{center}
\begin{list}{}{}
\item[${\mathrm{a}}$]{: the inner radius of the disk with a distance of 7.9 kpc and an inclination  angle of 29$^{\circ}$.}
\item[${\mathrm{b}}$]{: in unit of $10^{-2}$ photons/keV/cm$^2$/s at 1 keV.}
\end{list}
\end{table}


\clearpage

\begin{figure}[t]
\centering
   \includegraphics[angle=0, scale=0.4]{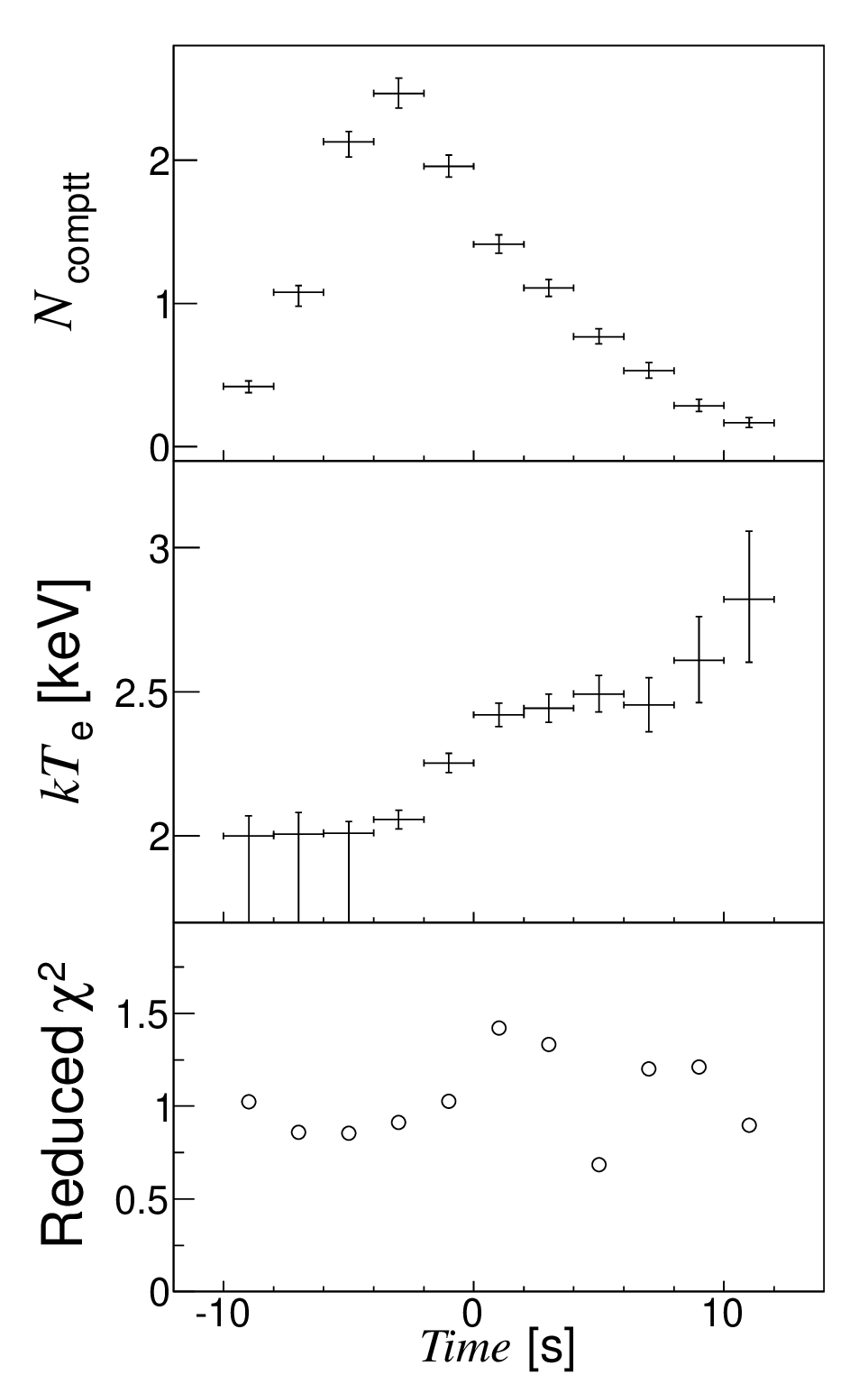}
   \includegraphics[angle=0, scale=0.4]{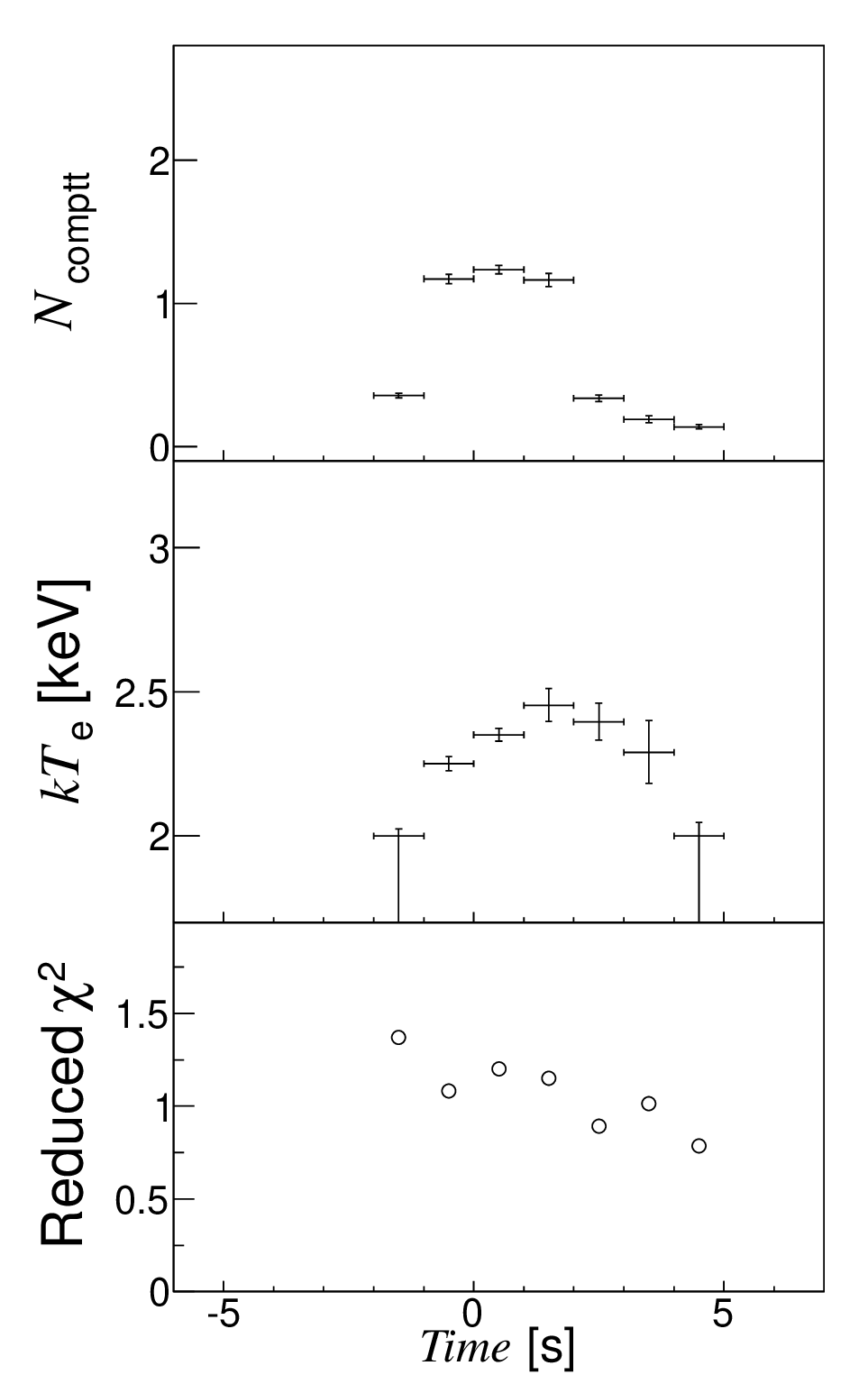}
 \caption{Time-resolved spectroscopy of the X-ray bursts using
an absorbed  comptt model for the TOO1 (the  left panel) and TOO2 (the  right panel), in  time bin 2 s and 1 s respectively.
The normalization $N_{\rm comp}$, the electron temperature $kT_{\rm e}$  and the reduced $\chi^{2}$ statistic are given in the top, middle and bottom panels respectively.
}
\label{burst_evolution}
\end{figure}


\end{document}